\documentstyle[preprint,tighten,aps]{revtex}

\def\l{\ell_{P}}

\def\SU{{\rm SU}}

\def\R{{\rm I}\!{\rm R}}
\def\E{\tilde{E}}
\def\A{\tilde{A}}

\preprint{\vbox{\baselineskip=12pt
\rightline{CGPG-98/3-2}}}

\begin{document}
\draft
\title{The Area Spectrum in Quantum Gravity}
\author{K.\ Krasnov\thanks{E-mail address: krasnov@phys.psu.edu}}
\address{Center for Gravitational Physics and Geometry, \\
The Pennsylvania State University, PA 16802, USA.}

\maketitle

\begin{abstract}
We show that, apart from the usual area operator of non-perturbative
quantum gravity, there exists another, closely related, operator
that measures areas of surfaces. Both corresponding classical 
expressions yield the area. Quantum mechanically, however, the
spectra of the two operators are different, coinciding only in 
the limit when the spins labelling the state are large. We argue that both
operators are legitimate quantum operators, and which one to
use depends on the context of a physical problem of interest. Thus,
for example, we argue that it is the operator proposed here 
that is relevant to use in the black hole context as measuring
the area of black hole horizon.
We show that the difference between the two operators is due to
non-commutativity that is known to arise in the quantum theory.
We give a heuristic picture explaining the difference
between the two area spectra in terms of quantum fluctuations 
of the surface whose area is being measured.

\end{abstract}

A well-known result of the loop approach \cite {Review} to quantum gravity 
is that the area of surfaces is quantized. In this theory
the spectrum of the operator
measuring the area of a surface $S$ is given by \cite{Area}
\begin{equation}
A_S=8\pi\gamma\l^2\sum_p\sqrt{j_p(j_p+1)},
\label{qarea}
\end{equation}
where the sum is taken over all points $p$ on the surface $S$ where edges of 
a spin network state transversely intersect $S$, $j_p$ are spins (half-integers) that 
label the intersecting edges, $\gamma$ is a real, positive parameter, known as
Immirzi parameter \cite{ImmPar}, and $\l^2=G\hbar$, $G$ being Newton constant.
It is assumed in (\ref{qarea}) that all intersections are transversal.

In this note we show that another quantity, 
that in the classical theory coincides with the area of $S$,
can be constructed,
but which gives rise to a quantum operator with a {\it different}
spectrum. We shall refer to this other
area operator and to its eigenvalues as $\A_S$. 
As we show below, the eigenvalues $\A_S$ are given by
\begin{equation}
\A_S = 8\pi\gamma\l^2\sum_p j_p,
\label{area}
\end{equation}
where the meaning of all the symbols is the same as in (\ref{qarea}). Thus,
$\A_S$ coincides with $A_S$ only in the limit all spins $j_p$ are
large. We shall argue that both $A_S$ and $\A_S$ give rise
to `legitimate' area operators, and a choice which of the area
operators should be used depends on the context of a physical
problem of interest.
We also give a heuristic interpretation of the
area spectra (\ref{qarea}), (\ref{area}) explaining the
discrepancy between the two. The area operator $\A_S$ was
also studied by Baez \cite{John}, in a similar context.

Before we give a description of the operator $\A_S$, let us recall
how the usual area operator \cite{Area} is constructed. Our choice of  
conventions is given in the Appendix. The area of a 2-surface
$S$ is given by 
\begin{equation}\label{1}
A_S = \int_S dx^1 \wedge dx^2 [{\rm Tr}(\E^3\E^3)]^{1/2},
\end{equation}
where we have used adapted coordinates such that $S$ is given 
by $x^3=0$. A simple way to understand the area spectrum 
(\ref{qarea}) is to rewrite the area operator as a sum
of angular momentum operators. In quantum theory the quantity $\E^{3i}$ gets
replaced by the operator of variational derivative 
$i\hbar 8\pi G\gamma (\delta/\delta A_{3i})$. The operator
$i(\delta/\delta A_{3i})$, when acting
on the holonomy along a spin net edge intersecting $S$,
behaves as the product of 
the 2-dimensional delta-function and the angular momentum operator
acting on the copy of the gauge group corresponding to the edge.
Introducing the angular momentum operators $X_p^i$ (one for each
point $p$ where $S$ is intersected with an edge of a spin network),
one can rewrite the area operator in the following simple form
\begin{equation}\label{6}
\hat{A}_S = 8\pi \l^2\gamma \sum_p [X_p^i X_{pi}]^{1/2},
\end{equation}
where the angular momentum operators satisfy the 
commutation relations of (times $1/2$) Pauli matrices
\begin{equation}\label{comm}
[X_p^i,X_p^j]=i\varepsilon^{ijk}X_p^k.
\end{equation}
Thus, each $X_p^i$ is the usual angular momentum
operator of quantum mechanics, which explains the spectrum
(\ref{qarea}). For details of this construction see \cite{Area}.

The expression (\ref{1}) giving the area of $S$ is a (gauge-invariant)
functional on the phase space of the theory. The other area operator
we are going to construct is {\it not} a quantization of a
gauge-invariant functional on the phase space. To construct it
we work in the space of non gauge-invariant
states. Let us fix a unit
vector $r^i$ (with values in the Lie algebra of $\SU(2)$) 
defined at each point on $S$. For an arbitrary unit vector $r^i:
r^i r_i =1$ let us consider the following quantity:
\begin{equation}\label{2}
\int_S dx^1 \wedge dx^2 \E^{3i} r_i.
\end{equation}
So far the quantity constructed has little to do with 
the area of $S$. Indeed, it is not gauge-invariant and not 
positive definite. However, one can extremize the quantity 
(\ref{2}) by performing a gauge rotation on the $\E^a$ field
and making it point in the same direction (in the internal space) 
as $r^i$ at each point on $S$.
Clearly, (\ref{2}) extremized in this way is just the area of $S$.
By $\A_S$ we will always mean the quantity (\ref{2}) where
$\E^{3i}$ is chosen to point in the same direction as $r^i$.

Although the above procedure gives the area of $S$ on the
classical level, it may seem to be very hard to perform
the `extremization' procedure in the quantum theory.
We shall see, however, that there is a natural analog
of this `extremization'.
It is not hard to construct an operator corresponding to (\ref{2}).
Indeed, one proceeds along the lines of the construction of 
the operator $\hat{A}_S$, replacing $i(\delta/\delta A_{3i})$ by 
the two-dimensional delta-function times the angular
momentum operator $X_p^i$. The corresponding
operator is given by
\begin{equation}\label{4}
8\pi \l^2\gamma \, \sum_p X_p^i r_i.
\end{equation}
Note that this operator is defined only in the space of 
non gauge-invariant states. It is easy to find its eigenvalues.
Denoting the eigenvalues
of the $X_p^i r_i$ operator by $m$ ($m$ are half-integers), we get
for the eigenvalues of the operator corresponding to (\ref{2})
\begin{equation}\label{3}
8\pi \l^2\gamma \sum_p m_p,
\end{equation}
where the eigenvalues $m_p: |m_p| \leq j_p$, $j_p$ being
the spins labelling the edges. Note that spin networks are
{\it not} eigenstates of (\ref{4}).

It is now easy to see what we have to do to recover the area
operator from (\ref{4}). Different eigenvalues (\ref{3}) 
have natural interpretation as different ways to project
$\E^{3i}$ vector on $r^i$ direction. Classically one gets the area
of $S$ when $\E^{3i}$ points in the same direction as $r^i$
and this is the maximal value for (\ref{2}) one can get.
The quantum mechanical analog of this is to take the maximal
eigenvalue of the operator corresponding to (\ref{2}).
It is natural to interpret this maximal eigenvalue 
as the eigenvalue of the area operator $\A_S$
\begin{equation}\label{5}
\A_S = 8\pi \l^2\gamma \sum_p j_p.
\end{equation}
Thus, as claimed, one can indeed construct an expression for the 
area different from (\ref{1}). Classically, both expression
give the area of $S$, quantum mechanically, however, the 
spectra of the corresponding operators are different,
coinciding only in the limit of large spins.

It may seem that the area operator $\A_S$ is pathological.
Indeed, it is defined using a quantity that is not 
gauge-invariant, and the construction involves the
procedure of extremization, i.e., of taking the maximal
eigenvalue, which may not be easy to make precise
in the quantum theory. However, there is one context
in which the area operator $\A_S$ is certainly legitimate.
It is the case of quantum theory on a manifold
with boundary. Dealing with such a theory one 
usually has to impose boundary conditions appropriate
to the physical problem of interest. These boundary
conditions may be such that they partially or
completely brake gauge-invariance on the boundary.
For example, in the construction of the black hole
sector of quantum theory in \cite{ABCK} a unit
vector $r^i$ has to be fixed on one of the boundaries
of space (black hole horizon), and the only gauge
transformations that survive on the
boundary are the ones fixing $r^i$. If this is
the case, the quantity (\ref{2}) constructed on 
the boundary is a gauge-invariant functional on the
phase space. The area $\A_S$ is a legitimate area
operator in this context, and it eigenvalues are
given by (\ref{5}).

In our opinion, both area operators discussed are of physical
interest, and which operator one has to use depends
on the context of a physical problem one deals with.
Thus, it would be good to have an interpretation
of the two area operators and to understand better
the discrepancy in their spectra.

We have given expressions (\ref{6}), (\ref{4})
for the two operators
in terms of angular momentum operators $X_p^i$.
Both operators have a
similar structure: they are both given by the sum over
the intersection points $p$. Thus, to understand the 
difference between them, we can concentrate on what happens
at a single intersection point. The contribution from
each point $p$ to (\ref{6}) is just the square root of
the operator $(X^i X_i)\equiv X^2$, the contribution to (\ref{4})
is the projection of $X^i$ on the direction $r^i$,
where $X^i$ is the usual angular momentum operator.
Thus, as we know from the theory of angular momentum,
the difference between the spectra (\ref{qarea}), (\ref{area})
arises because the components of the angular momentum
do not commute (\ref{comm}). This is why the
eigenvalue of $X^2$ is given by $j(j+1)$, not by $j^2$.
One can understand the discrepancy by
using a heuristic idea of uncertainty that arises
because of the non-commutativity of the components of
$X^i$. Let us consider a representation of highest
weight $j$, and let us visualize the angular momentum
in this representation as a vector in $\R^3$. 
The length of this vector is $\sqrt{j(j+1)}$.
The vector $r^i$ gives a preferred direction in this internal space,
and the maximal eigenvalue of $(X\cdot r)$ operator in this
representation is $j$. This can be visualized as the
maximal projection of $X^i$ vector on $r^i$ direction 
allowed by the uncertainty relations. Indeed, the 
vector $X^i$ can not point solely in the direction $r^i$
because this would mean that the other (orthogonal to $r^i$)
components of the vector are zero. However, if at least one
components of $X^i$ is non-zero, all
three components of the vector $X^i$ cannot have definite
values simultaneously, due to their non-commutativity. 
A maximal possible value of the projection of $X^i$
on $r^i$ direction must be accompanied by non-zero
(and undetermined) values of the other two components
of $X^i$, such that $(X^{\bot})^2:=X^2-(X\cdot r)^2$
is equal to $j$. This can be visualized as vector
$X^i$ precessing (fluctuating) about $r^i$ direction,
with neither of the two components of $X^{\bot}$ having
definite values, but satisfying $(X^{\bot})^2=j$.

Thus, we get the following interpretation of the
discrepancy between the two area operators.
At each intersection point $p$ of the surface
$S$ with an edge of a spin network, one has the 
quantum angular momentum vector $X_p^i$. The area
operator (\ref{6}) measures the length $\sqrt{j_p(j_p+1)}$
of this vector, while the other area
operator (\ref{4}) measures the maximal possible
projection $j_p$ of $X_p^i$ on $r^i$ direction.
The two differ because of the non-commutativity
(and associated uncertainty) of the components of
vector $X_p^i$. Note that the non-commutativity
we have discussed is closely related to the 
non-commutativity of area operators discussed
in \cite{Zapata}.

This picture can be further visualized by choosing an 
identification of the internal space $\R^3$ at
each point of $S$ with the tangent space to the manifold. This
is equivalent to choosing some (non-physical)
triad field. Let us choose an identification
in such a way that 2-flats orthogonal to the
image of $r^i$ are integrable and span the surface
$S$. Thus, we will visualize $r^i$ as the vector
normal to the surface $S$. Then $r^i (X_p\cdot r)$ is
a vector at $p$ orthogonal to $S$, whose length
is $j_p$. We can associate with it a 2-flat at $p$,
with the area $8\pi\l^2\gamma j_p$. Then the
area (\ref{area}) of $S$ is simply the sum
of contributions coming from all such 2-flats.
The vector $X_p^i$ has the length $\sqrt{(j_p(j_p+1)}$,
and we can associate with it a 2-flat at $p$
of the area $8\pi\l^2\gamma\sqrt{j_p(j_p+1)}$.
However, in the visual picture this vector
(and the corresponding two-flat) is constantly fluctuating.
The collection of 2-flats corresponding to $X_p^i$
does no longer span the surface $S$, but can be visualized
as $S$ fluctuating. The area of this fluctuating surface
is given by the sum of the areas of the 2-flats, i.e.,
by (\ref{qarea}).

Thus, the interpretation of the two areas 
(\ref{qarea}), (\ref{area}) we propose is that
quantum mechanically there are two surfaces
instead of one classical, and the two different areas 
are areas of these two surfaces.
When defined intrinsically in terms of 
quantum geometry described by $\E^a$ operator
(as a collection of the 2-flats orthogonal to $X_p^i$),
the surface $S$ is fluctuating 
due to quantum mechanical uncertainties,
as we sketched above. Without
a vector $r^i$ this is the only surface
we have, and its
area is given by (\ref{qarea}). However, if one
chooses a vector $r^i$ and defines $S$
as being spanned by the 2-flats orthogonal to
$r^i$, the surface is not fluctuating, and
its area is given by (\ref{area}). The two 
areas coincide in the limit of large spins
(relative small fluctuations).

Concluding, we would like to repeat that in our
opinion both area operators studied in this note
are physically interesting. It is the context
of a problem that should decide which operator
should be used to measure the area of a surface.
Here we gave a heuristic interpretation
of the two areas as those of fluctuating and non-fluctuating
surfaces. This interpretation, however,
should be considered as only suggestive.
More work is necessary to make it precise.

The last comment is that, in our opinion, it is the area (\ref{area})
that should be used to measure the area of black 
hole horizon in the approach of \cite{ABCK}. In this approach the
black hole horizon is treated as an internal spacetime
boundary, where a special set of boundary conditions
is imposed. One of this boundary conditions is 
exactly the requirement that there is a fixed 
unit vector $r^i$ on the spatial cross-section of the horizon.
Moreover, the boundary conditions also fix several components
of the triad field $\E^a$ on the horizon: 1,2-components in 
the coordinate system used in (\ref{1}). These components
of the triad field are not subject to quantization, and can thus be
used to identify the internal space at each point of $S$ with
the tangent space, as above. With this identification
the surface $S$ can be defined as 
spanned by the 2-flats orthogonal to $r^i$. 
As we sketched above, the boundary surface defined with respect
to $r^i$ is not fluctuating. Thus, one has to use
the formula (\ref{area}) for the area eigenvalues. The
usage of the area spectrum (\ref{area}) instead of (\ref{qarea})
resolves the following puzzle in the  approach \cite{ABCK}
noted by Carlip \cite{Carlip}. 
The quantity $A/8\pi\l^2\gamma$, where $A$ is the horizon
area, plays the role of the level of the
associated Chern-Simons theory (see \cite{ABCK} for 
details). However, the level is required to be an
integer, which is not the case if the area spectrum
is given by (\ref{qarea}). If, on the other hand, one uses the 
spectrum (\ref{area}), which we argued is the right
spectrum to be used in the context of black holes,
the level is given by the sum of spins labelling
the black hole state, which is 
an integer due to the requirement of gauge invariance. 

The following important point should be made regarding
our proposal to use the area spectrum (\ref{area})
as relevant in the context of black holes.
The spectrum (\ref{area}) is equidistant, i.e.,
the separation between two adjacent eigenvalues
is constant, and equal to $4\pi\l^2\gamma$.
Thus, this area spectrum effectively reduces
to the one proposed in 1974 by Bekenstein \cite{Bek}.
However, as it was recently shown by Bekenstein
and Mukhanov \cite{BM}, such area spectrum leads
to the black hole emission spectrum qualitatively
different from the one predicted by the 
semi-classical calculation of Hawking \cite{Haw}.
More precisely, the spectrum proposed by Bekenstein
leads to the purely discrete emission spectrum,
consisting of distinct, equidistant emission lines,
while the semi-classical spectrum is continuous.
However, as it was shown first in \cite{BHspec}, the 
spectrum (\ref{qarea}) is {\it not} equidistant, 
the separation between two adjacent lines 
for large horizon areas $A$ goes as $1/\sqrt{A}$.
Thus, if the black hole is described by (\ref{qarea}),
although the area spectrum is discrete, the separation
between eigenvalues is very small for large areas,
and the emission spectrum effectively reproduces 
continuous spectrum. This may lead one to conclude that
(\ref{qarea}), not (\ref{area}), is the correct spectrum to describe the
quantum black hole. However, we have argued that the
difference between the two spectra (\ref{qarea}),
(\ref{area}) is due to quantum fluctuations of the
surface whose area is being measured, - in this
case the quantum event horizon. This suggests that,
in the case the spectrum (\ref{area}) is used,
the distinct lines in this spectrum get `smeared'
out by quantum fluctuations. Thus, the spectrum (\ref{area})
will lead to qualitatively the same black hole
emission spectrum as (\ref{qarea}). The picture
is similar to the one considered by M\"akel\"a
\cite{Makela}. Thus, the usage of
the spectrum (\ref{area}) {\it does
not} seem to lead to any undesirable properties of the
black hole emission spectrum. 

This can be made more precise using the analysis of 
the black hole emission spectrum given in \cite{Rad}.
Considering quantum transitions in which the spin labelling
one of the edges changes, one obtains the
spectrum consisting of distinct emission lines, with the 
enveloping curve being thermal. One finds that the 
separation between lines 
in the emission spectrum is $\Delta(\hbar\omega/2\pi T)=\Delta(j)=1/2$,
where $\omega$ is the frequency of the quantum emitted and
$T$ is the black hole temperature. As we have discussed 
above, the contribution to the horizon area from each
edge labelled with spin $j$ should be thought of as
fluctuating, with the amplitude of fluctuations being
$\sqrt{j(j+1)}-j$. In the black hole transition
processes the main role is played by transitions
with the final state being $j=1/2$. Thus, taking $j=1/2$,
we get for the fluctuation amplitude $\sqrt{3}/2-(1/2)\approx0.37$.
The width $D$ of the emission line is proportional to
twice the fluctuation amplitude $D(\hbar\omega/2\pi T)\approx 2*0.37$,
which is larger then the separation 
$\Delta(\hbar\omega/2\pi T)$ between the lines. Thus, the 
emission spectrum is effectively continuous.

The above argument is tentative. A more
detailed study is necessary to understand all
implications of the presence of the two area spectra
in the theory. Note that
an additional motivation for the usage of  
spectrum (\ref{area}) comes from the study of
rotating black holes, in particular extremal
ones. As it is shown in \cite{Essay},
it is spectrum (\ref{area}) that is compatible with
certain desirable properties of the angular
momentum in the theory.

Another implication of our results in the context of
black holes is that the numerical value of the 
proportionality coefficient between the black hole 
entropy and the area changes. In the approach \cite{ABCK},
the black hole entropy turns out to be given by $N\ln{2}$,
where $N$ is the number of transverse intersections of
spin $1/2$ that is needed to give the total horizon area of 
black hole $A$. Using the area spectrum (\ref{area}) instead
of (\ref{qarea}), we get $N=A/4\pi\l^2\gamma$. This implies
that the entropy is given by
\begin{equation} \label{entropy}
S = {\gamma_0\over\gamma} {A\over 4\l^2}, \qquad \qquad 
\gamma_0:={\ln{2}\over\pi} \approx 0.22.
\end{equation}
This is similar in the form to the result obtained in \cite{ABCK},
the only difference being the numerical value of coefficient
$\gamma_0$. The entropy still cannot be compared to
Bekenstein-Hawking formula because of the presence of the free
parameter $\gamma$ in (\ref{entropy}).

\acknowledgements

I am grateful to Jose Zapata for a discussion, and to 
John Baez for suggestions on the first version of the
manuscript. This work was supported 
in part by the Braddock Fellowship of the Pennsylvania State
University, NSF grant PHY95-14240 and by the Eberly research funds 
of the Pennsylvania State University.

\appendix
\section{Conventions}

The phase space of the theory is that of $\SU(2)$ Yang-Mills: the 
configurational variable is described by an $\SU(2)$ connection $A_a$, 
and its conjugated momentum is $\E^a$, which can be thought of as
a Lie algebra valued field. The momentum $\E^a$
has the geometrical interpretation of an orthonormal triad field
with density weight one
\begin{equation}
\tilde{\tilde{g}}^{ab} = - {\rm Tr}(\E^a\E^b).
\end{equation}
All traces in this paper are traces in the fundamental representation
of $\SU(2)$. The density weight of $\E$ is indicated by the single 
`tilde' over the symbol. It is convenient to introduce
the components of fields $A_a, \E^a$ with respect to
a basis in the Lie algebra. We take 
\begin{eqnarray} 
A_a:= -{i\over2}\tau^i A_a^i \\
\E^a:= -{i\over\sqrt{2}}\tau^i \E^{ai},
\nonumber
\end{eqnarray}
which are the standard conventions in the literature. 
Here $\tau^i$ are the usual Pauli matrices 
$(\tau^i\tau^j)=i\varepsilon^{ijk}\tau^k +
\delta^{ij}$. The Poisson brackets between
the canonical fields are given by
\begin{equation}\label{poisson}
\{A_{ai}(x),\E^{bj}(y)\} = 8\pi G\gamma \delta_a^b \delta_i^j 
\tilde\delta^3(x,y),
\end{equation}
where $\delta_a^b, \delta_i^j$ are Kronecker deltas, and $\tilde\delta^3$
is Dirac's delta-function (densitized) in three dimensions. Note that
the connection variable we use is the real connection 
$A=\Gamma_a-\gamma K$, where $\gamma$ is the Immirzi parameter and
$\Gamma, K$ are the spin connection and the extrinsic curvature
correspondingly, while $\E^a$ is the usual densitized triad field.
This explains the appearance of $\gamma$ in the Poisson brackets
(\ref{poisson}).

\end{document}